\documentclass[runningheads]{llncs}

\usepackage{tabularx}

\usepackage{soul}
\usepackage{url}
\usepackage{hyperref}
\usepackage[utf8]{inputenc}
\usepackage[small]{caption}
\usepackage{graphicx}
\usepackage{amsmath}
\usepackage{booktabs}
\usepackage[table]{xcolor}% http://ctan.org/pkg/xcolor
\definecolor{grey}{rgb}{0.8,0.8,0.8}
\usepackage{algorithm}
\usepackage{algorithmicx}
\usepackage{algpseudocode}
\usepackage{amsmath,amssymb}
\usepackage{mathtools}

\DeclarePairedDelimiterX{\infdivx}[2]{(}{)}{%
  #1\;\delimsize\|\;#2%
}
\newcommand{\infdiv}{D_{KL}\infdivx}
\DeclarePairedDelimiter{\norm}{\lVert}{\rVert}

\newcommand{\joint}{p \left( z_{t+1}, u_t \middle| z_t \right)}
\newcommand{\marginaltransition}{p \left( z_{t+1} \middle| z_t \right)}

\newcommand{\planning}{p \left( u_{t} \middle| z_t, z_{t+1} \right)}
\newcommand{\source}{\omega \left( u_{t} \middle| z_t \right)}
\newcommand{\variational}{q \left( u_{t} \middle| z_{t+1}, z_t \right)}
\newcommand{\gradient}{\frac{\partial}{\partial\theta}}

\usepackage{xcolor}

\newcommand{\robot}{~SCR~}

\usepackage[numbers]{natbib}
\usepackage{adjustbox}
\usepackage{pgfplots}
\usepgfplotslibrary{statistics}
\usepackage{scrextend}
\usepackage{siunitx}
\usepackage[acronym, toc, shortcuts, nowarn]{glossaries}
\glsdisablehyper
\setacronymstyle{long-short}
\newacronym{RL}{RL}{Reinforcement Learning}
\newacronym{IRL}{IRL}{Inverse Reinforcement Learning}
\newacronym{DL}{DL}{Deep Learning}
\newacronym{DNN}{DNN}{Deep Neural Network}
\newacronym{DRL}{DRL}{Deep Reinforcement Learning}
\newacronym{IL}{IL}{Imitation Learning}
\newacronym{IGP}{IGP}{Interacting Gaussian Process}
\newacronym{RVO}{RVO}{Reciprocal Velocity Obstacles}
\newacronym{ORCA}{ORCA}{Optimal Reciprocal Collision Avoidance}
\newacronym{SFM}{SFM}{Social Force Model}
\newacronym{TDM}{TDM}{Temporal Difference Method}
\newacronym{SCR}{SCR}{Socially Compliant Robot}
\newacronym{CADRL}{CADRL}{Collision Avoidance with Deep Reinforcement Learning}
\newacronym{LSTM-RL}{LSTM-RL}{Long Short Term Memory - Reinforcement Learning}
\newacronym{SARL}{SARL}{Socially Attentive Reinforcement Learning}
\newacronym{IB}{IB}{Invisible Baseline}

\begin{document}

\title{Social Navigation with Human Empowerment driven Deep Reinforcement Learning}

\author{Tessa van der Heiden\inst{1} \and
Florian Mirus\inst{1} \and
Herke van Hoof\inst{2}}

\authorrunning{T. van der Heiden et al.}
\institute{BMW Group, Landshuter Str. 26, 85716 Unterschleißheim
\email{tessa.heiden@bmw.de} \email{florian.mirus@bmwgroup.de}\\ \and
University of Amsterdam, Science Park, 1098 XH Amsterdam\\
\email{h.c.vanhoof@uva.nl}}
\maketitle              % typeset the header of the contribution
\begin{abstract}
Mobile robot navigation has seen extensive research in the last decades.
The aspect of collaboration with robots and humans sharing workspaces will become increasingly important in the future. 
Therefore, the next generation of mobile robots needs to be socially-compliant to be accepted by their human collaborators. 
However, a formal definition of compliance is not straightforward.
On the other hand, empowerment has been used by artificial agents to learn complicated and generalized actions and also has been shown to be a good model for biological behaviors. 
In this paper, we go beyond the approach of classical \acf{RL} and provide our agent with intrinsic motivation using empowerment.
In contrast to self-empowerment, a robot employing our approach strives for the empowerment of people in its environment, so they are not disturbed by the robot's presence and motion.
In our experiments, we show that our approach has a positive influence on humans, as it minimizes its distance to humans and thus decreases human travel time while moving efficiently towards its own goal. 
An interactive user-study shows that our method is considered more social than other state-of-the-art approaches by the participants.

\end{abstract}


\section{Introduction}
\begin{figure}[t!]
  \centering
    \includegraphics[trim={4.5cm 3.5cm 6.5cm 10.5cm},clip, width=.7\linewidth]{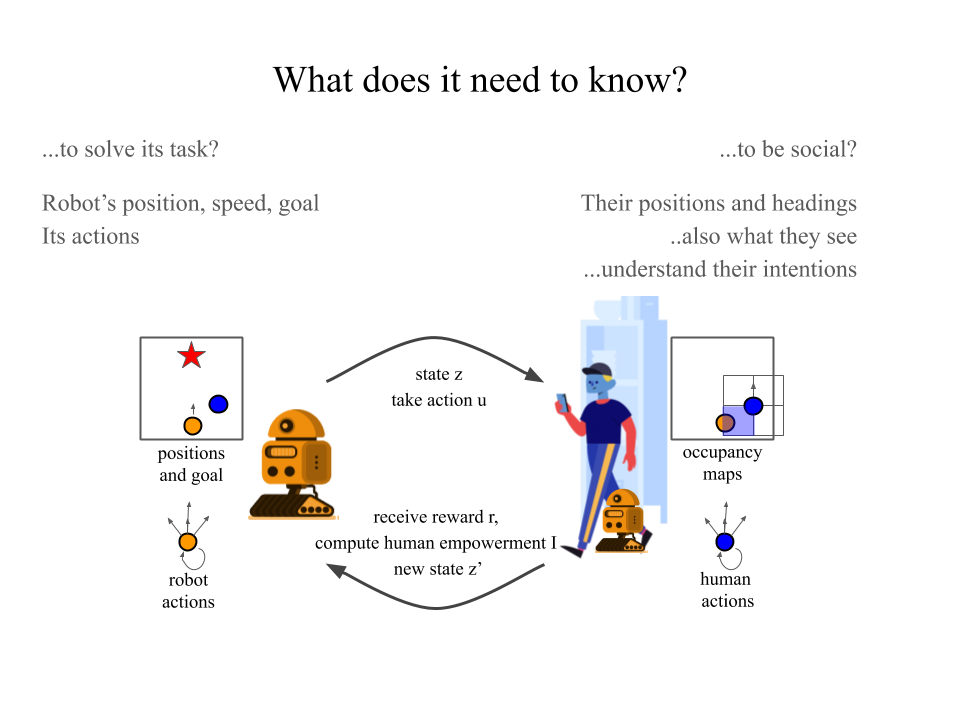}
  \caption{Our social compliant robot (SCR) uses occupancy maps centered around each human to compute human empowerment, so it minimally disturbs people to pursue their goals. }
  \label{fig: max_path}
  \vspace{-0.5cm}
\end{figure}

Recent advances in sensor and control technologies have allowed the development of robots assisting people in domestic, industrial, and traffic environments.
One key challenge in such settings, where humans and robots share the same workspace, is that the robot must plan safe, collision-free paths, which have to be socially compliant for the humans to accept robots as collaborators in the long run.

However, moving and thereby interacting with people requires robots to follow specific unwritten social rules, for instance, politely keeping their distance \cite{templeton2018walking}, which depends not only on the situation \cite{sieben2017collective} but also the social context and the people involved \cite{robicquet2016learning}. According to \citet{kruse2013human}, the three main requirements for a robot to navigate in a socially compliant way are \textit{comfort}, \textit{naturalness}, and \textit{sociabilit}y. Robots acting according to these rules will have a higher chance of acceptance by human users.

% What methods tackle this?
Existing methods for social robot navigation either model social conventions between agents explicitly \cite{pfeiffer2016predicting}, or implicitly learn them through \ac{IL} \cite{shiarlis2017acquiring}, or even through \acf{RL} \cite{chen2017socially}.
However, explicitly defining rules or reward functions for social navigation is not straightforward, while generating a sufficiently large body of training examples for imitation learning can be cumbersome or infeasible.

Empowerment \cite{klyubin2015empowerment} allows the agent to generate rewards by itself and offers a useful alternative.
It is the channel capacity between actions and future states and maximizes an agent's influence on its near future.
In contrast to self-empowerment, where a self-empowered agent will try to push others away to maximize its future rewards, an agent who strives for others' empowerment maintains the influence of them on their futures \cite{salge2017empowerment}.

In this paper, we propose a novel approach to social robot navigation employing a combination of \ac{RL} and human empowerment to provide a robot with intrinsic motivation. 
An agent employing our approach strives for people's empowerment to minimize disturbance when pursuing their goals and respect people's personal space.
Our contribution is to use the concept of human empowerment introduced by \citet{salge2017empowerment} as an intrinsic reward function for \ac{RL} for social navigation.

In an extensive evaluation in a simulation environment, we compare our method with state-of-the-art robotic navigation methods.
Inspired by \cite{kruse2013human}, we use two additional metrics, the distance between human and robot and robot's jerk, assessing social behavior to evaluate our approach.
Additionally, we study the robot's influence on people and vice-versa by introducing two new metrics, the travel time and distance of humans and the robot. Last, we assess our approach in a user-study.  Our experiments show that our approach can achieve socially compliant robot navigation by using human empowerment. 
Finally, our method applies to any multi-agent system that requires a robot to interact with humans in a socially compliant way, since it does not require a cost function.

\section{Related work}
Many approaches have designed interaction models to enhance social awareness in robot navigation.
We discuss these methods first and motivate the practicality of \ac{DRL}.
We proceed by describing empowerment as a member of a family of intrinsic motivators for  \ac{RL}.

\subsection{Social navigation}
The goals for social navigation can be divided into three main categories: \textit{comfort}, \textit{naturalness} and \textit{sociability} \cite{kruse2013human}.
Examples of \textit{comfort} are respecting personal space, avoiding erratic behavior, and not interfering with the other's movement.
\textit{Naturalness} is mostly related to how similar a robot's motion is to human behavior, e.g., smooth and interpretable, while \textit{sociability} is mainly associated with social conventions and etiquettes.
Previous works have tried to create navigation frameworks satisfying all of those requirements.
Well-engineered methods are the \ac{SFM} \cite{helbing1995social}, \ac{IGP} \cite{trautman2010unfreezing}, \ac{ORCA} \cite{karamouzas2009predictive} and \ac{RVO} \cite{van2008reciprocal}.
Although all of these methods yield collision-free paths, they rely on manually engineered models with limited additional social characteristics.

% Deep learning approaches?
In contrast to such model-based approaches, \ac{DL} models have shown to produce more human-like paths~\cite{gu2014humanlike}.
For instance, \acp{DNN} allow policies to better comply with humans' social rules \cite{cross2019social}.
Early works separate the prediction of the environment and the policy's planning task into two neural networks \cite{bansal2019learningcontrol}, which could cause the freezing robot problem since the predicted human motion could occupy all the available future space \cite{trautman2010unfreezing}. 

% Reinforcement learning
\acf{IL} and \ac{IRL} obtain policies directly from demonstrations \cite{pfeiffer2016predicting}, which requires an extensive data set due to the uncertainty of human motion.
Alternatively, \ac{DRL} aims to learn cooperative strategies by interacting with the environment \cite{chen2018crowd}.
However, the definition of a suitable cost function that encourages a robot to navigate socially is a challenging task.
Even if a cost function might appear evident in some cases (e.g., collision-free and keeping distance to neighbors), it often has to be regularised to achieve smooth and risk-averse behavior.

\subsection{Empowerment}
% What is empowerment and where is it applied?
Instead of shaping the reward function to achieve the desired behavior, an emerging field within \ac{RL} focuses on intrinsic motivation \cite{oudeyer2007intrinsic}.
There are many different ways to motivate an agent intrinsically, and one possible technique is called empowerment \cite{klyubin2005empowerment}, \cite{salge2014empowerment}.
Empowerment was applied to teach agents task-independent behavior and training in settings with sparse rewards, such as stabilizing an inverted pendulum, learning a biped to walk \cite{karl2017unsupervised}. \citet{aubret2019survey} provides a comprehensive survey of empowerment for \ac{RL}.

Earlier approaches were only applicable in discrete state-action spaces, but recently \cite{mohamed2015variational} show efficient implementations for continuous settings.
In our work, we will build upon these models.

\section{Methodology}
% What is the task of the robot?
Our goal is to teach an agent on how to navigate its target in a socially compliant manner safely. A combination of two rewards can achieve these two objectives. In this section, we describe our agent and the two types of rewards.

% What is the problem formulation?
We consider the system to be Markovian, where each next state $x_{t+1}$ depends only on the current state $x_t$ and agent's action $u_t$ and no prior history. A value network model  $V_\phi$ is trained to accurately approximate the optimal value function $V^*$ that implicitly encodes social cooperation between agents and the empowerment of the people, see Equation \ref{eq: v}. 
\begin{equation}
\begin{aligned}
V^*(x_t) &= \sum_{t=0}^T \gamma^t R_t(x_{t+1}, \pi^*(x_t)) 
%R_t(z_{t+1}, \pi^*(z_t)) &= I_t(z_{k, t+1},u_{k,t}^\omega|z_{k,t}) + \beta \cdot R_{t, e}(z_t, u_t^\pi)
\end{aligned}
\label{eq: v}
\end{equation}
$R_t(\cdot)$ is the reward function and $\pi^*$ is the optimal policy that maximizes the expected return, with discount factor $\gamma\in(0,1)$. 
\subsection{Reward for safe navigation}
The first task of the agent is to reach its goal while avoiding collisions and keeping a comfortable distance to humans. We consider the number of humans in each episode to be fixed.

% How is this goal reached?
Equation \ref{eq: rewards} defines the environmental reward function $R_{t, e}(x_t,  u_t^\pi)$ for this task with the robot's state denoted as $x_t$ and its action with $u_t^\pi$. Similar to other \ac{DRL} methods for social navigation \cite{chen2017decentralized}, \cite{chen2018crowd}, \cite{chen2017socially} we award task accomplishments and penalize collisions or uncomfortable distances.
\begin{align}
R_{t, e}(x_t, u_t^\pi)&{}= 
\begin{cases}
-.25  				& \quad \text{if } d_i < 0.01 \\
-0.1$+$  \frac{d_i}{2}  & \quad \text{else if } 0.01 \leq d_i \le 0.2 \\
1                       & \quad \text{else if }  d_g \leq 0.01 \\
0                       & \quad \text{otherwise}
\end{cases}
\label{eq: rewards}
\end{align}
Here $d_g = \norm{\overline{p} -\overline{p}_g}_2$ is the robot’s distance to the goal during a time interval $\Delta t$ and $d_i= \norm{\overline{p} -\overline{p}_i}_2$ is the robot’s distance to neighbor $i$. It gets rewarded when its current position $\overline{p}$ reaches the position of the goal $\overline{p}_g$, but penalized if its position is too close to another one's position $\overline{p}_i$.  

% What do we need to know?
The robot's own state, $\mathbf{x}$, consists of a 2D position vector $\overline{p}=\left[p_x, p_y\right]$ and 2D velocity vector $\overline{v}=\left[v_x, v_y\right]$. The human states are denoted by $\mathbf{X} = \left[X_1, X_2,..., X_k\right]$, which is a concatenated vector of states of all $k$ humans participating in the scene. Each entry is similar to the robot's state, namely, $X_i = \left[\overline{p}_{i}, \overline{v}_{i}\right]$. The final state of the robot is the concatenation of the state of the humans and robot, $x_t=[\mathbf{X}_t, \mathbf{x}_t]$. Its action is a desired velocity vector, $u_t^\pi=\overline{v}_d$

\subsection{Empowerment for social compliance}
The robot's second task is to consider people in its neighborhood and respond to their intentions in a socially compliant manner. Designing a suitable reward function is a challenging task, among other things, due to the stochasticity in people's behaviors. This is where we use empowerment \cite{klyubin2005empowerment}, \cite{salge2014empowerment}, an information-theoretic formulation of an agent’s influence on its near future. 

\subsubsection{Human empowerment}
% Why human empowerment % Why do we expect it to give us social traits?
Empowerment in our case, motivates the robot to approach states in which its neighbors are most empowered. Now the robot aims to maximize the empowerment of another person rather than its own, which \citet{salge2017empowerment} call \textit{human empowerment} in contrast to \textit{robot empowerment}.
As a result, the robot will prevent obstructing the human, for example, by getting too close or by interfering with the human's actions, both of which \citet{kruse2013human} defined as social skills. 

%On the contrary, \citet{jaques2019social} have a completely different approach for multi-agent reinforcement learning (MARL). In their work, agents obtain an intrinsic reward for having a causal influence on other agents’ actions, while our policy is rewarded for giving maximal decision freedom to others by minimally disturbing them.
% What is empowerment?
Equation \ref{eq: empowerment} describes the definition of empowerment $\varepsilon$ being the maximal mutual information $I$ for a state $z_t$ \cite{klyubin2005empowerment}. It is the channel capacity between action $u_t^{\omega}$ and future state $z_{t+1}$, maximized over source policy $\omega$. Policy $\omega$ is part of the human’s decision making system.%\footnote{To reduce notation clutter, we substitute $z_t$ and $z_{t+1}$ with $z$ and $z'$ and drop superscript $\omega$ when they can be inferred from the context. }. 
\begin{equation}
\begin{aligned}
\varepsilon(z_t)
&=\max_\omega I (z_{t+1}, u_t^\omega | z_t)
%=\max_\omega \underbrace{H(u|z)}_{\text{a}} - \underbrace{H(u|z, z')}_{\text{b}} 
= \max_\omega H(u_t^\omega|z_t) - H(u_t^\omega|z_t, z_{t+1}) 
\end{aligned}
\label{eq: empowerment}
\end{equation}
% What does empowerment do practically?
The right part defines the empowerment with entropies $H(\cdot)$. It corresponds to increasing the diversity of decisions, while at the same time limiting those decision that have no effect. Intuitively, the empowerment of the person reflects his or her ability to influence their future.

%$\pi$ is trained to take an action $u_r$ in each state such that the expected empowerment value $\epsilon(z_k)$ of the next state is highest.
The human state $z_t$ takes an ego-centric parameterization \cite{chen2017decentralized}. Each state is an occupancy grid map centered around the person, denoted with $\mathbf{g}_k$. It is a 3D tensor with dimensions $c \times r \times 3$, where $c$ and $r$ run over the height and width of the grid. Each entry contains the presence and velocity vector of a neighbor $j$ at that location $\overline{e}_j = [1, v_x, v_y]$. The resulting state of the humans is a concatenated vector denoted by $z_t = \left[\mathbf{g}_1, \mathbf{g}_2,..., \mathbf{g}_k\right]$ and action are continuous values in $\mathbb{R}^2$. 

\subsubsection{Estimating empowerment with neural networks}
To compute empowerment, we consider the mutual information defined by the Kullback-Leibler divergence \cite{kullback1951information} between the joint $\joint$ and the product of the marginal distributions $\marginaltransition$ and $\source$: 
\begin{equation}
\begin{aligned}
   I &= \infdiv{\joint}{\marginaltransition \source}  \\
   &= \iint \joint  \ln{\frac{\joint}{\marginaltransition \source}} dz' du
\end{aligned}
\label{eq: KL}
\end{equation}
% How do other deal with it?
The main problem in the formulation in Eq. \ref{eq: KL} is the intractability due to the integral of all future states. Since the introduction of empowerment \cite{klyubin2005empowerment}, \cite{salge2014empowerment} many have designed methods to deal with this. Recent works provide an efficient method to estimate a lower bound on empowerment $\hat{I}$, via variational methods \cite{karl2017unsupervised}, \cite{mohamed2015variational}, \cite{burda2015importance}. 
\begin{equation}
\begin{aligned}
   \hat{I} = \iint \joint  \ln{\frac{\variational}{ \source}} dz' du
\end{aligned}
\end{equation}
Instead of $\marginaltransition$ a planning distribution $\planning$ is used, which is approximated with the variational approximation $\variational$ to obtain a lower bound. $\hat{I}$ can now be maximized over the parameters of the source, $\source$ and variational $\variational$ networks.  $\planning$ is a third neural network that computes the future state $z_{t+1}$ from $z_t$ and $u_t^\omega$

The gradient can be computed as follows, in which the joint parameters of $\omega(\cdot)$ and $q(\cdot)$ is denoted by $\theta$:
\begin{equation}
\begin{aligned}
   \gradient \hat{I} = \gradient \mathbb E_{\joint} \left[\ln \frac{\variational}{\source}\right]
\end{aligned}
\end{equation}
Using Monte-Carlo integration to estimate the continuous case, we can obtain the following gradient:
\begin{equation}
\begin{aligned}
    \gradient \hat{I}
   \approx \sum_{n=1}^{N} \gradient \left[\ln\left(\variational\right) - \ln( \source) \right]
\end{aligned}
\end{equation}
%The source policy $\omega$ is part of the human’s decision making system, which is a hypothetical distribution that does not take any actions in the external environment and will be used to estimate $I$. 
We are free to choose any type of distribution and since human movement is not discrete, we model both $\variational = \mathcal{N}(\mu_{q},\,\sigma_{q}^{2})$, $p(z'|u,z) = \mathcal{N}(\mu_{p},\,\sigma_{p}^{2})$ and $\source = \mathcal{N}(\mu_{\omega},\,\sigma_{\omega}^{2})$ as Gaussian distributions. 

\subsection{Training procedure}
The robot with policy $\pi$ learns to safely navigate to its goal and achieve human empowered states. This is achieved by training a value network $V_\phi$  with the reward function combining the mutual information $\hat{I}_t(\cdot)$ and the environmental reward $R_{t,e}(\cdot)$. The hyper-parameter $\beta$ is used to regulate the trade-off between social compliance and safety:
\begin{equation}
    R_t(z_{t}, x_{t}, u_t^\omega, u_t^\pi) = (1-\beta)\cdot I_t(z_{t+1},u_{t}^\omega|z_{t}) + \beta \cdot R_{t, e}(x_t, u_t^\pi)
\end{equation}

%Algorithm \ref{alg:v-learning} shows the full algorithm. 
A set of demonstrations from the \ac{ORCA} policy is used to give the robot a head start. \ac{ORCA} describes a deterministic control policy \cite{van2008reciprocal} and its demonstrations speed up learning, because experiences in which the robot reaches the goal are now part of the memory. 
Next, the behavior policy $\pi$ collects samples of experience tuples  $e_t = (z_t, x_t, u_t^\pi, r_{t, e}, z_{t+1}, x_{t+1})$ until a final state is reached. Random actions are selected with probability $\epsilon$. Once these are collected, our hypothetical human policy $\omega$ together with $q$ and $p$ are used to estimate $\hat{I}_t$. Finally, the networks are trained with a random mini-batch obtained from the memory ($e^{b}$). 
The value network $V_\phi$ is optimized by the \ac{TDM} \cite{sutton1998introduction} with standard experience replay and fixed target network techniques \cite{chen2017decentralized}, \cite{mnih2015human},  \cite{chen2018crowd}. 

\begin{equation}
\begin{aligned}
y_t &= r_{e,t}^{(b)} + \hat{I}_{t}^{(b)} + \gamma^t  \hat{V}_{\hat{\phi}}(z_{t+1}^{(b)}, x_{t+1}^{(b)}) \\
    \phi &\leftarrow \phi +  \lambda \nabla_\phi  (y_t-V_\phi(z_t, x_t))^2 
\end{aligned}
\end{equation}

$\hat{V}_{\hat{\phi}}$ denotes the target network. The networks $\omega_\theta$ and $q_\theta$ updated through gradient ascent and $p_\psi$ via gradient descent:

\begin{equation}
\begin{aligned}
     \theta &\leftarrow \theta +  \lambda \nabla_\theta \hat{I}_t \\
      \psi &\leftarrow \psi +  \lambda \nabla_\psi (z_{t+1}^{(b)}-p(u_t, z_t))^2 
\end{aligned}
\end{equation}
The behavior policy $\pi$ uses the joined state $x_t$ to navigate collision-free to its goal. $p$, $q$ and $\omega$ take the occupancy grids centered around each human $z_t$ as states for the computation of $\hat{I}_t$.

\section{Experiments}
We conduct three experiments to evaluate our proposed model. The first experiment compares our model against four existing state-of-the-art methods. The second experiment assess the social competences, based on the metrics defined by \citet{kruse2013human}. The final experiment consists human subjects that evaluate the models in an interactive simulator. 
%Lastly, we investigate the effectiveness of our method through qualitative analysis.

\subsection{Implementation details}
The simulator used in this work is from \cite{chen2018crowd}. It starts and terminates an episode with five humans and the robot. The human's decisions are simulated by \citet{van2011rvo2}, which uses the \ac{ORCA} policy \cite{van2008reciprocal} to calculate their actions. \ac{ORCA} uses the optimal reciprocal assumption to avoid other agents. In \ref{se:human_ev}, another simulator is used to control the position of one human manually and terminates once the human reaches its goal. The code and videos can be found online\footnote{\label{note1}\url{https://github.com/tessavdheiden/SCR}}.

We implemented the networks in PyTorch and trained them with a batch size of \num{100} for \num{10000} episodes. For the value network, the learning rate is $\lambda_{v} = 0.001$ and the discount factor $\gamma$ is \num{0.9}. The exploration rate of the $\epsilon$ decays linearly from \num{0.5} to \num{0.1} in the first \num{5000} episodes and stays \num{0.1} for the remaining \num{5000} episodes. These values are the same as \citet{chen2018crowd}.

The parameter $\beta$ is \num{0.25}, because that gave the highest discomfort distance rate and success rate. The learning rates for the other networks are similar. The value network is trained with stochastic gradient descent, identical to \citet{chen2018crowd}. The planning, source, and transition networks are trained with Adam \cite{kingma2014adam}, similar to \cite{karl2017unsupervised}. 

\subsection{State-of-the-art navigation benchmark}
Table \ref{tab: benchmark} reports the rates of success, collision, the
robot navigation time, discomfort distance, and the average discounted
cumulative reward averaged over \num{500} episodes. 
\textbf{Success} is the rate of robot reaching its goal within a time limit of \SI{20}{\second}. \textbf{Collision} is the rate of the robot colliding with humans. \textbf{Disc. dist.} is the rate in which the distance between the robot and a human was smaller than $0.1$. 
We compare our robot with four existing state-of-the-art methods, \ac{ORCA} \cite{van2011rvo2}, \ac{CADRL} \cite{chen2017decentralized},  \ac{LSTM-RL} \cite{everett2018motion} and \ac{SARL} \cite{chen2018crowd}. As can be seen, our \ac{SCR} and SARL both outperform other baselines on the standard metrics. Next, we look more thoroughly into the performance of \ac{SCR} and \ac{SARL}.
\begin{table}
 \vspace{-.5cm}
\caption{Both \ac{SCR} and \ac{SARL} outperform the other baselines, which can be seen by the best values (grey). \ac{ORCA} does not have any collisions, because this is the central idea behind the method (*). The numbers are computed for 500 different test scenarios.}
\label{tab: benchmark}
\begin{center}
%\begin{adjustbox}{max width=\linewidth}
%\resizebox{\columnwidth}{!}{
\begin{tabular}{l *5c}
\toprule
     &   \textbf{Success}     & \textbf{Collision}                 & \textbf{Robot }            	& \textbf{Disc.}                  & \textbf{Reward}         \\ &  \textbf{rate \% }     & \textbf{rate \%}                & \textbf{time}              	& \textbf{dist. \%}                  &          \\
     \midrule
\ac{ORCA}        &   0.99               & .000*                     & 12.3              & 0.00*                     & .284          \\ 
\ac{CADRL}       &   0.94               & .035                      & 10.8              & 0.10                      & .291                  \\ 
\ac{LSTM-RL}     &   0.98               & .022                      & 11.3             	& 0.05                      & .299                     \\ 
\ac{SARL}        &  \cellcolor{grey} 0.99   & .002  & \cellcolor{grey} 10.6		& \cellcolor{grey}0.03     & \cellcolor{grey}.334 \\ 
\ac{SCR} (ours) &  \cellcolor{grey}  0.99   & \cellcolor{grey}.001                  & 10.9                      & \cellcolor{grey}0.03     & .331          \\\bottomrule
\end{tabular}
%
%\end{adjustbox}
\end{center}
  \vspace{-1.cm}
\end{table}

\subsection{Influence of robot on humans and vice-versa}
Table \ref{tab: path length} shows travel times and distances of both humans and the robot. \textbf{Robot time} and  \textbf{Human time} are the average navigation times of the agents to reach their goals. Keeping both low indicates that the robot does not disturb humans and moves quickly to its target. The path length of the robot, \textbf{Trav. distance}, is calculated to make sure that it moves without unnecessary detours. The simulator allows making the robot invisible to the humans, which is called the Invisible baseline (\ac{IB}). This baseline serves as a testbed for validating the other policies' abilities in reasoning about the interactions with humans. 

\ac{IB} has the highest travel time and distance because it does not influence humans. On the contrary, \ac{SARL} has a low travel distance and time, but human travel times are highest. \ac{SARL} has learned that humans avoid it. The travel times of \ac{SCR} and that of the humans are nearly the same. These numbers suggest that our method has learned to minimally disturb other people while moving to its own goal efficiently due to human empowerment. 

\begin{table}
\caption{\ac{SCR} moves efficiently to its target and doesn't disturb people as their travel times are low. \ac{SARL} has learned that people avoid it, so the humans' travel times are higher than its own. The \ac{IB} takes a considerable detour because people do not see it. 
 Moreover, \robot has the lowest jerk ($\frac{m}{s^3}$), since it avoids being close to a person and non-smooth behavior as this would lower the empowerment of its neighbors.}
\label{tab: path length}
\begin{center}
%\begin{adjustbox}{max width=\linewidth}
%\resizebox{\columnwidth}{!}{
\begin{tabular}{l *5c}
\toprule
 & \textbf{Robot}   			& \textbf{Trav}.  	& \textbf{Human}   & \textbf{Sep}.   			& \textbf{Jerk} 					\\
 &  \textbf{time}  			& \textbf{distance} 	& \textbf{time}   & \textbf{distance}  			& 					\\\midrule
\ac{IB}          & 11.5                          &  10.7                 & \cellcolor{grey}9.1      &\cellcolor{grey}.85 & .73   \\                  
SARL                        & \cellcolor{grey}10.6    	    &  \cellcolor{grey}9.2  & 10.7                     &.41                 & .77    \\  
\ac{SCR} (ours)               & 10.9                          &  9.3  			    & \cellcolor{grey}9.1      & .43                &\cellcolor{grey} .51   \\ \bottomrule
\end{tabular}
%}
%\end{adjustbox}
\end{center}
  \vspace{-0.8cm}
\end{table}

Next, we examine how we can evaluate social compliance further. \citet{fong2003survey} and \citet{kruse2013human} state that people judge robots negatively if the separation distance between them is low and move non-smoothly. \textbf{Sep. distance} is the distance between a human and the robot and \textbf{Jerk} is the jerk of the robot ($\frac{m}{s^3}$).

The reported numbers are in the last two columns of Table \ref{tab: path length}. Even though on average \ac{SCR} and \ac{SARL} are close to humans, they do not exceed a minimum distance of .1m, see collision rate in Table \ref{tab: benchmark}. 
The \ac{IB} and \ac{SARL} move with a high jerk because their reward functions do not incorporate it. On the other hand, \ac{SCR} has the lowest jerk because it avoids erratic behavior as that would lower humans' empowerment. 

\subsection{Human evaluation}
\label{se:human_ev} 
A successful navigation strategy can best be tested with real persons. To that end, 30 persons interacted with the robot and controlled the position of one human in the simulator \footref{note1}. 
The robot and humans start at \SI{90}{\degree} from each other and need to cross the center to reach their goals. The simulator terminates once the human reaches his/her goal. After that, subjects are asked to rate the social performance of the robot with a score from \num{1} to \num{10} (similar to the study performed by \cite{shiarlis2017acquiring}).  A one-way repeated-measure analysis of variance (ANOVA) was conducted to determine significance, and a significant effect was obtained ($f(2, 27)=13.485$, $p<.01$). 

\begin{figure}
\centering
  \begin{minipage}[b]{0.35\linewidth}
  \begin{tikzpicture}
\begin{axis}[
ylabel = \textbf{Human Score},
width=1.5\linewidth,
boxplot/draw direction=y,
x axis line style={opacity=0},
axis x line*=bottom,
axis y line=left,
enlarge y limits,
enlarge x limits=0.5,
ymajorgrids,
xtick={1,2,3},
xticklabels={IB, SARL, SCR},
ymin = 0, ymax=10,
ytick={0, 5, 10}
]
\addplot+[
boxplot prepared={
lower whisker=2, lower quartile=3,
median=5,
upper quartile=6.5, upper whisker=8,
},
]
%table[row sep=\\,y index=0] { 0\\ };
coordinates {};
\addplot+[
boxplot prepared={
lower whisker=2, lower quartile=4,
median=5,
upper quartile=6, upper whisker=7,
},
]
coordinates {(0,35) (0,55)};
\addplot+[
boxplot prepared={
lower whisker=6, lower quartile=6.5,
median=7.5,
upper quartile=8.5, upper whisker=9,
},
]
coordinates {};
%table[row sep=\\,y index=0] { 9\\ };
\end{axis}
\end{tikzpicture}

%http://www.chiark.greenend.org.uk/doc/texlive-doc/latex/pgfplots/pgfplots.pdf
  \end{minipage}
    \hspace{2.00cm}
  \caption{ The box plots summarize human evaluation scores on three methods, the \ac{IB}, \ac{SARL} and our method (\ac{SARL}). Both the \ac{IB} and \ac{SARL} have almost the same median, but for the \ac{IB} the humans' scores deviate more from the median. Our method obtains the highest score.}
  \label{fig: interactive}
    \vspace{-0.5cm}
\end{figure}
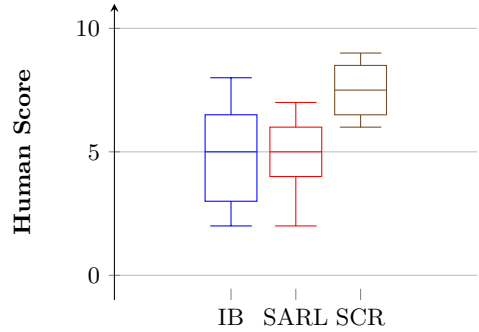

As can be seen in Fig. \ref{fig: interactive}, both \ac{SARL} and the \ac{IB} have similar medians, but the samples of the baseline deviate more from the median.  The \ac{IB} started to move away from the human, even before the human started to move. \ac{SARL}, on the other hand, moves directly to its own goal. \ac{SCR} hast the highest score, which shows the potential of our method. 

\subsection{Qualitative results}
\begin{figure}
 \vspace{-.5cm}
  \centering
  \begin{minipage}[b]{0.24\linewidth}
    \includegraphics[trim={7cm 2.5cm 6.5cm 2cm},clip, width=\linewidth]{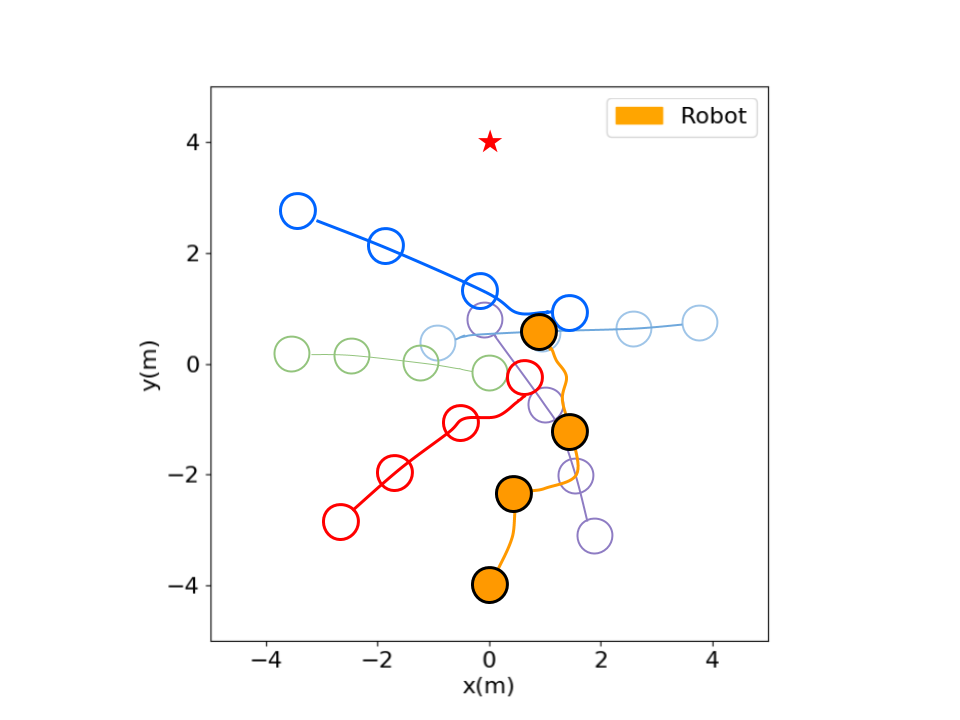}
   \centering (a) \ac{SARL} t=6
  \end{minipage}
  \hfill
  \begin{minipage}[b]{0.24\linewidth}
    \includegraphics[trim={7cm 2.5cm 6.5cm 2cm},clip, width=\linewidth]{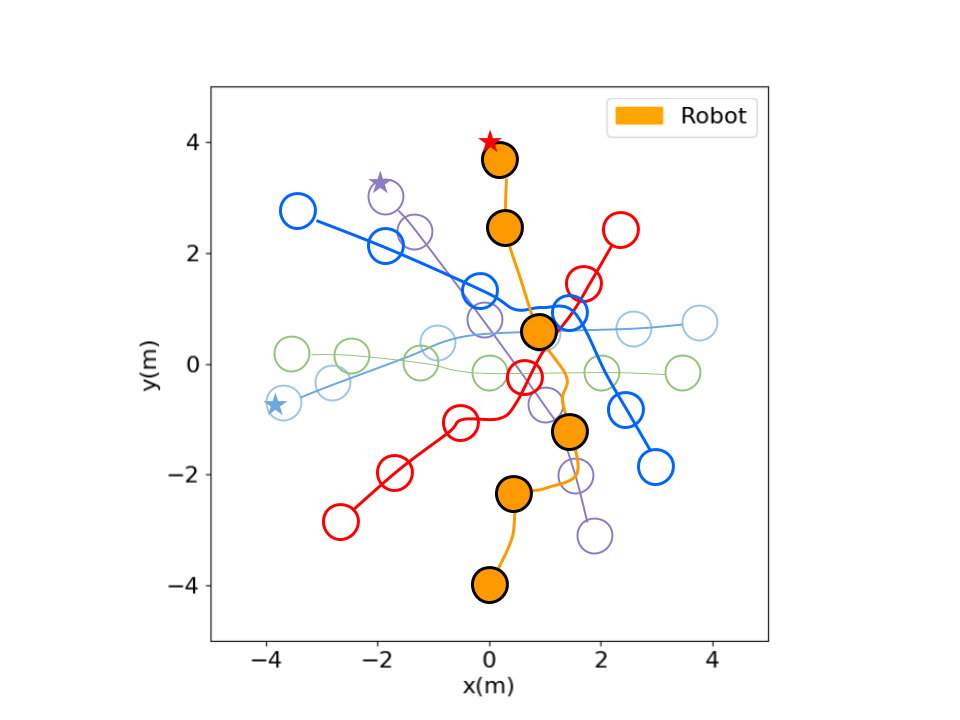}
    \centering (b) \ac{SARL} t=9.2
  \end{minipage}
      \hfill
    \begin{minipage}[b]{0.24\linewidth}
    \includegraphics[trim={7cm 2.5cm 6.5cm 2cm},clip, width=\linewidth]{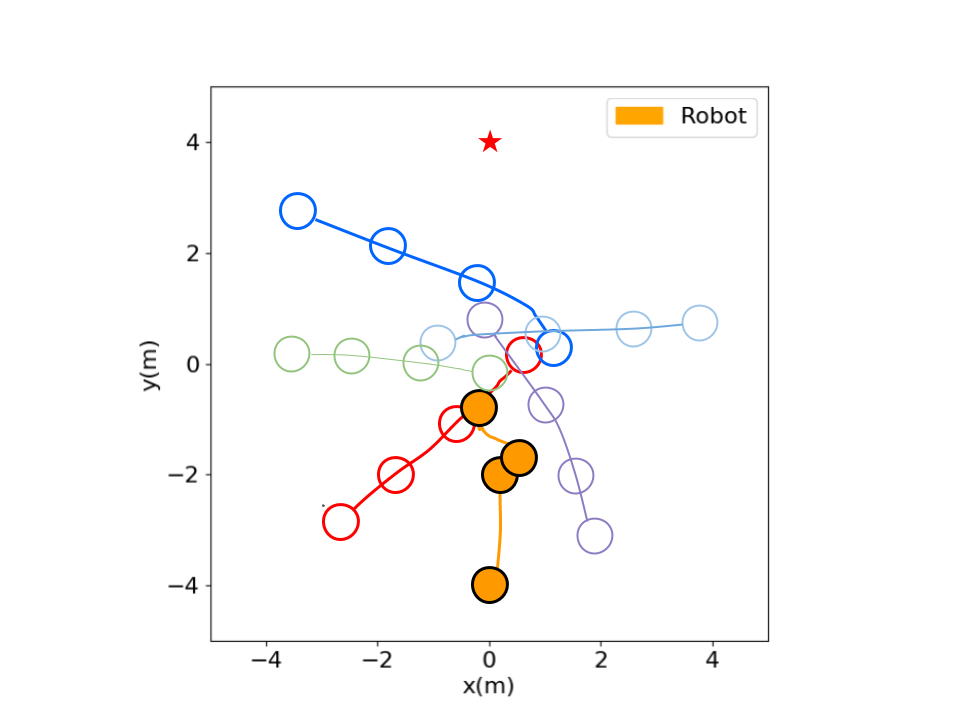}
   \centering (c) \ac{SCR} t=6
  \end{minipage}
          \hfill
    \begin{minipage}[b]{0.24\linewidth}
    \includegraphics[trim={7cm 2.5cm 6.5cm 2cm},clip, width=\linewidth]{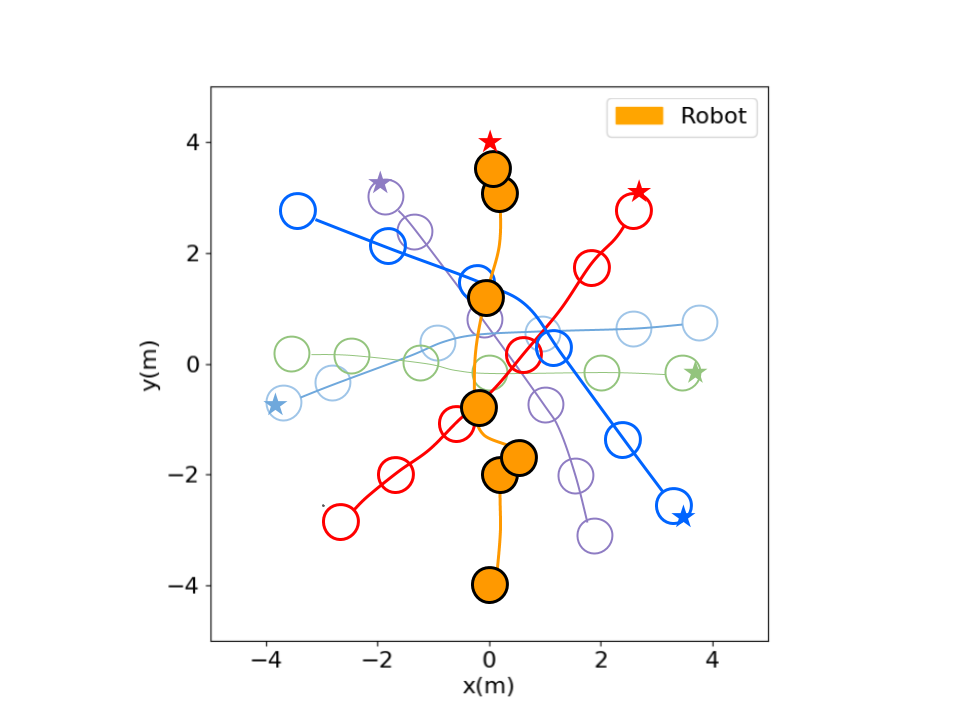}
   \centering (d) \ac{SCR} t=10.5
  \end{minipage}
  \caption{\ac{SARL} (a, b) and \ac{SCR} (c, e) in a scene with 5 humans. The humans' destinations are the opposite of the x, y-axis' origin from their initial locations. \ac{SARL} reaches its destination quickly, but only two out of five humans reach it (b, 2 stars). Two persons (red, blue) need to adjust their path to avoid the robot (orange). \ac{SCR} waits at t=6 (c) and all humans reach their destination (d, 5 stars). }
  \label{fig: im_max_path}
   \vspace{-1.cm}
\end{figure}

Figure \ref{fig: im_max_path} shows \ac{SARL} and \ac{SCR} navigating through a crowd of five people. The left figures shows \ac{SARL} (a, b) and right \ac{SCR} (c, d) at two different time steps.  The trajectories indicate that SARL goes directly to its goal, while \ac{SCR} waits at $t=6$ (c). Moreover, at $t=9.2$, SARL has reached its goal, but only two out of five humans reach theirs (b, purple and light blue stars). In contrast, \ac{SCR} reaches its goal at t=10.5, but all people reached their final destinations (d). \ac{SARL} overtakes two people (a, red and green) and alters the path of another (a, blue). On the contrary, \ac{SCR} lets them pass (c, red, green and blue). SARL uses occupancy maps to model the pairwise interaction between humans \cite{chen2018crowd}, so it cannot incorporate the robot's influence on each human. On the contrary, \ac{SCR} uses empowerment maps for each human with high values in states in which it does not block anyone. 

% How to compute f-stat: http://sphweb.bumc.bu.edu/otlt/MPH-Modules/BS/BS704_HypothesisTesting-ANOVA/BS704_HypothesisTesting-Anova_print.html
% Calulate https://goodcalculators.com/one-way-anova-calculator/
% Critical values: https://www.itl.nist.gov/div898/handbook/eda/section3/eda3673.htm

\section{Conclusion and future work}
This paper proposed a reinforcement learning method for social navigation with intrinsic motivation using the empowerment of surrounding people. Our approach avoids the hard-coded reward signals and allows people nearby not to be disturbed by the robot.
Our experiments show that our policy outperforms other methods on social characteristics.
The influence of the robot's motion is difficult to evaluate by people in simulation. Thus, we also compared the methods in an interactive simulator and obtained positive results. 
For future work, we would like to extend the model to deal with a variable amount of humans and with different policies. It would also be interesting to extend the method to incorporate the effect that (non-moving) objects have on humans.

\bibliographystyle{named}
\bibliography{root}  

\begin{thebibliography}{}

\bibitem[\protect\citeauthoryear{Aubret \bgroup \em et al.\egroup
  }{2019}]{aubret2019survey}
A.~Aubret, L.~Matignon, and S.~Hassas.
\newblock A survey on intrinsic motivation in reinforcement learning.
\newblock {\em arXiv preprint arXiv:1908.06976}, 2019.

\bibitem[\protect\citeauthoryear{Bansal \bgroup \em et al.\egroup
  }{2019}]{bansal2019learningcontrol}
S~Bansal, V~Tolani, S~Gupta, J~Malik, and C~Tomlin.
\newblock Combining optimal control and learning for visual navigation in novel
  environments.
\newblock {\em arXiv preprint arXiv:1903.02531}, 2019.

\bibitem[\protect\citeauthoryear{Burda \bgroup \em et al.\egroup
  }{2015}]{burda2015importance}
Y.~Burda, R.~Grosse, and R.~Salakhutdinov.
\newblock Importance weighted autoencoders.
\newblock {\em arXiv preprint arXiv:1509.00519}, 2015.

\bibitem[\protect\citeauthoryear{Chen \bgroup \em et al.\egroup
  }{2017a}]{chen2017socially}
Y.~Chen, M.~Everett, M.~Liu, and J.~P. How.
\newblock Socially aware motion planning with deep reinforcement learning.
\newblock In {\em 2017 IEEE/RSJ IROS}, pages 1343--1350. IEEE, 2017.

\bibitem[\protect\citeauthoryear{Chen \bgroup \em et al.\egroup
  }{2017b}]{chen2017decentralized}
Y.~Chen, M.~Liu, M.~Everett, and J.~P. How.
\newblock Decentralized non-communicating multiagent collision avoidance with
  deep reinforcement learning.
\newblock In {\em 2017 IEEE ICRA}, pages 285--292. IEEE, 2017.

\bibitem[\protect\citeauthoryear{Chen \bgroup \em et al.\egroup
  }{2019}]{chen2018crowd}
C.~Chen, Y.~Liu, S.~Kreiss, and A.~Alahi.
\newblock Crowd-robot interaction: Crowd-aware robot navigation with
  attention-based deep reinforcement learning.
\newblock In {\em 2019 ICRA}, pages 6015--6022. IEEE, 2019.

\bibitem[\protect\citeauthoryear{Cross \bgroup \em et al.\egroup
  }{2019}]{cross2019social}
E.~Cross, R.~Hortensius, and A.~Wykowska.
\newblock From social brains to social robots: applying neurocognitive insights
  to human-robot interaction.
\newblock {\em Philosophical Transactions of the Royal Society of London.
  Series B, Biological sciences}, 374, 04 2019.

\bibitem[\protect\citeauthoryear{Everett \bgroup \em et al.\egroup
  }{2018}]{everett2018motion}
M.~Everett, Y.~Chen, and J.P. How.
\newblock Motion planning among dynamic, decision-making agents with deep
  reinforcement learning.
\newblock In {\em 2018 IEEE/RSJ IROS}, pages 3052--3059. IEEE, 2018.

\bibitem[\protect\citeauthoryear{Fong \bgroup \em et al.\egroup
  }{2003}]{fong2003survey}
T.~Fong, I.~Nourbakhsh, and K.~Dautenhahn.
\newblock A survey of socially interactive robots.
\newblock {\em Robotics and autonomous systems}, 42(3-4):143--166, 2003.

\bibitem[\protect\citeauthoryear{Gu and Dolan}{2014}]{gu2014humanlike}
T.~Gu and J.~Dolan.
\newblock Toward human-like motion planning in urban environments.
\newblock In {\em 2014 IEEE Intelligent Vehicles Symposium Proceedings}, pages
  350--355. IEEE, 2014.

\bibitem[\protect\citeauthoryear{Helbing and Molnar}{1995}]{helbing1995social}
D.~Helbing and P.~Molnar.
\newblock Social force model for pedestrian dynamics.
\newblock {\em Physical review E}, 51(5):4282, 1995.

\bibitem[\protect\citeauthoryear{Karamouzas \bgroup \em et al.\egroup
  }{2009}]{karamouzas2009predictive}
I~Karamouzas, P~Heil, P.~Van~Beek, and M.H. Overmars.
\newblock A predictive collision avoidance model for pedestrian simulation.
\newblock In {\em International workshop on motion in games}, pages 41--52.
  Springer, 2009.

\bibitem[\protect\citeauthoryear{Karl \bgroup \em et al.\egroup
  }{2017}]{karl2017unsupervised}
M.~Karl, M.~Soelch, P.~Becker-Ehmck, D.~Benbouzid, P.~van~der Smagt, and
  J.~Bayer.
\newblock Unsupervised real-time control through variational empowerment.
\newblock {\em arXiv preprint arXiv:1710.05101}, 2017.

\bibitem[\protect\citeauthoryear{Kingma and Ba}{2014}]{kingma2014adam}
D.P. Kingma and J.~Ba.
\newblock Adam: A method for stochastic optimization.
\newblock {\em arXiv preprint arXiv:1412.6980}, 2014.

\bibitem[\protect\citeauthoryear{Klyubin \bgroup \em et al.\egroup
  }{2005a}]{klyubin2005empowerment}
A.S. Klyubin, D.~Polani, and C.L. Nehaniv.
\newblock Empowerment: A universal agent-centric measure of control.
\newblock In {\em 2005 IEEE Congress on Evolutionary Computation}, volume~1,
  pages 128--135. IEEE, 2005.

\bibitem[\protect\citeauthoryear{Klyubin \bgroup \em et al.\egroup
  }{2005b}]{klyubin2015empowerment}
A.S. Klyubin, Daniel Polani, and Chrystopher Nehaniv.
\newblock Empowerment: A universal agent-centric measure of control.
\newblock volume~1, pages 128 -- 135 Vol.1, 10 2005.

\bibitem[\protect\citeauthoryear{Kruse \bgroup \em et al.\egroup
  }{2013}]{kruse2013human}
T.~Kruse, A.K. Pandey, R.~Alami, and A.~Kirsch.
\newblock Human-aware robot navigation: A survey.
\newblock {\em Robotics and Autonomous Systems}, 61(12):1726--1743, 2013.

\bibitem[\protect\citeauthoryear{Kullback and
  Leibler}{1951}]{kullback1951information}
Solomon Kullback and Richard~A Leibler.
\newblock On information and sufficiency.
\newblock {\em The annals of mathematical statistics}, 22(1):79--86, 1951.

\bibitem[\protect\citeauthoryear{Mnih \bgroup \em et al.\egroup
  }{2015}]{mnih2015human}
V.~Mnih, K.~Kavukcuoglu, D.~Silver, A.A. Rusu, J.~Veness, M.G. Bellemare,
  A.~Graves, M.~Riedmiller, A.K. Fidjeland, G.~Ostrovski, et~al.
\newblock Human-level control through deep reinforcement learning.
\newblock {\em Nature}, 518(7540):529, 2015.

\bibitem[\protect\citeauthoryear{Mohamed and
  Rezende}{2015}]{mohamed2015variational}
S.~Mohamed and D.J. Rezende.
\newblock Variational information maximisation for intrinsically motivated
  reinforcement learning.
\newblock In {\em NeurIPS}, pages 2125--2133, 2015.

\bibitem[\protect\citeauthoryear{Oudeyer \bgroup \em et al.\egroup
  }{2007}]{oudeyer2007intrinsic}
P~Oudeyer, F.~Kaplan, and V.V. Hafner.
\newblock Intrinsic motivation systems for autonomous mental development.
\newblock {\em IEEE Transactions on Evolutionary Computation}, 11(2):265--286,
  2007.

\bibitem[\protect\citeauthoryear{Pfeiffer \bgroup \em et al.\egroup
  }{2016}]{pfeiffer2016predicting}
M.~Pfeiffer, U.~Schwesinger, H.~Sommer, E.~Galceran, and R.~Siegwart.
\newblock Predicting actions to act predictably: Cooperative partial motion
  planning with maximum entropy models.
\newblock In {\em 2016 IEEE/RSJ IROS}, pages 2096--2101. IEEE, 2016.

\bibitem[\protect\citeauthoryear{Robicquet \bgroup \em et al.\egroup
  }{2016}]{robicquet2016learning}
A.~Robicquet, A.~Sadeghian, A.~Alahi, and S.~Savarese.
\newblock Learning social etiquette: Human trajectory understanding in crowded
  scenes.
\newblock In {\em European conference on computer vision}, pages 549--565.
  Springer, 2016.

\bibitem[\protect\citeauthoryear{Salge and Polani}{2017}]{salge2017empowerment}
C.~Salge and D.~Polani.
\newblock Empowerment as replacement for the three laws of robotics.
\newblock {\em Frontiers in Robotics and AI}, 4:25, 2017.

\bibitem[\protect\citeauthoryear{Salge \bgroup \em et al.\egroup
  }{2014}]{salge2014empowerment}
C.~Salge, C.~Glackin, and D.~Polani.
\newblock Empowerment--an introduction.
\newblock In {\em Guided Self-Organization: Inception}, pages 67--114.
  Springer, 2014.

\bibitem[\protect\citeauthoryear{Shiarlis \bgroup \em et al.\egroup
  }{2017}]{shiarlis2017acquiring}
Kyriacos Shiarlis, Joao Messias, and Shimon Whiteson.
\newblock Acquiring social interaction behaviours for telepresence robots via
  deep learning from demonstration.
\newblock In {\em 2017 IEEE/RSJ International Conference on Intelligent Robots
  and Systems (IROS)}, pages 37--42. IEEE, 2017.

\bibitem[\protect\citeauthoryear{Sieben \bgroup \em et al.\egroup
  }{2017}]{sieben2017collective}
A.~Sieben, J.~Schumann, and A.~Seyfried.
\newblock Collective phenomena in crowds—where pedestrian dynamics need
  social psychology.
\newblock {\em PLoS one}, 12(6):e0177328, 2017.

\bibitem[\protect\citeauthoryear{Sutton \bgroup \em et al.\egroup
  }{1998}]{sutton1998introduction}
R.S. Sutton, A.G. Barto, et~al.
\newblock {\em Introduction to reinforcement learning}, volume~2.
\newblock MIT press Cambridge, 1998.

\bibitem[\protect\citeauthoryear{Templeton \bgroup \em et al.\egroup
  }{2018}]{templeton2018walking}
A.~Templeton, J.~Drury, and A.~Philippides.
\newblock Walking together: behavioural signatures of psychological crowds.
\newblock {\em Royal Society open science}, 5(7):180172, 2018.

\bibitem[\protect\citeauthoryear{Trautman and
  Krause}{2010}]{trautman2010unfreezing}
P.~Trautman and A.~Krause.
\newblock Unfreezing the robot: Navigation in dense, interacting crowds.
\newblock In {\em 2010 IEEE/RSJ IROS}, pages 797--803. IEEE, 2010.

\bibitem[\protect\citeauthoryear{van~den Berg \bgroup \em et al.\egroup
  }{}]{van2011rvo2}
J.~van~den Berg, S.J. Guy, J.~Snape, M.C. Lin, and D.~Manocha.
\newblock Rvo2 library: Reciprocal collision avoidance for real-time
  multi-agent simulation.

\bibitem[\protect\citeauthoryear{Van~den Berg \bgroup \em et al.\egroup
  }{2008}]{van2008reciprocal}
J.~Van~den Berg, M.~Lin, and D.~Manocha.
\newblock Reciprocal velocity obstacles for real-time multi-agent navigation.
\newblock In {\em 2008 IEEE International Conference on Robotics and
  Automation}, pages 1928--1935. IEEE, 2008.

\end{thebibliography}

\end{document}